\documentclass[%
 aip,
 jmp,%
 amsmath,amssymb,
 reprint,%
]{revtex4-1}

\usepackage{graphicx}
\usepackage{dcolumn}
\usepackage{bm}

\begin{document}

\preprint{AIP/123-QED}

\title[Nonlinear Density Waves]{Nonlinear Density Waves in the Single-Wave Model\footnote{Error!}}

\author{Kiril B. Marinov}
 \altaffiliation{ASTeC, STFC Daresbury Laboratory and The Cockcroft Institute, Keckwick Lane, Daresbury,
 WA4 4AD, United Kingdom.}

\author{Stephan I. Tzenov}
 \altaffiliation{Department of Physics, Lancaster University and The Cockcroft Institute, Keckwick Lane, Daresbury,
 WA4 4AD, United Kingdom.}
\email{s.tzenov@lancs.ac.uk}


\date{\today}

\begin{abstract}
The single-wave model equations are transformed to an exact hydrodynamic closure by using a class of solutions to the Vlasov equation corresponding to the waterbag model. The warm fluid dynamic equations are then manipulated by means of the renormalization group method. As a result, amplitude equations for the slowly varying wave amplitudes are derived. Since the characteristic equation for waves has in general three roots, two cases are examined. If all three roots of the characteristic equation are real, the amplitude equations for the eigenmodes represent a system of three coupled nonlinear equations. In the case, where the dispersion equation possesses one real and two complex conjugate roots, the amplitude equations take the form of two coupled equations with complex coefficients. The analytical results are then compared to the exact system dynamics obtained by solving the hydrodynamic equations numerically.
\end{abstract}

\maketitle

%

\section{\label{sec:intro}Introduction}

The processes of pattern and coherent structures formation in plasmas and plasma-like media have attracted attention for many years. A number of approximations and simplified models has been proposed, amongst which the single-wave model is one of the most efficient approaches to study the weakly nonlinear behavior in marginally stable plasmas. Starting from the general Vlasov-Maxwell equations for the phase space density distribution and the self-consistent electromagnetic field, one usually derives a coupled set of equations describing the evolution of the coarse-grained distribution function and a certain isolated marginally stable wave mode of the electrostatic potential.

Using the method of matched asymptotic expansions, the single-wave model equations have been recently derived \cite{DelCastillo} in the most general case independent of the equilibrium. A rigorous derivation of the nonlinear evolution of a unstable electrostatic wave for a multispecies Vlasov plasma has been given by Crawford and Jayaraman \cite{Crawford} and this general physical picture has been shown to correspond to the single-wave model. Amongst earlier pioneering work utilizing the single-wave model, the studies dedicated  to the beam-plasma instability \cite{Drummond,Oneil,Matsiborko} and the bump-on-tail instability \cite{Simon,Janssen,Denavit} should be mentioned. The influence of the finiteness of the number of particles coupled to a monochromatic wave in a collisionless plasma has been investigated by M-C. Firpo et al \cite{Firpo,Firpo1}. Based on the utilization of the single-wave picture some analogies between two classical models of the single-pass free electron laser dynamics and of the beam-wave plasma instability have been discussed by A. Antoniazzi et al \cite{Antoniazzi}. This implies that the results presented in the present paper will be useful to describe processes of formation of patterns and coherent structures in single-pass free electron lasers.

The purpose of the present paper is not to argue on the validity or the justification of the single-wave picture, rather than to use it as a starting point for deriving amplitude equations for the slowly varying wave envelopes. It is organized as follows. In the next section, we cast the self-consistent single-wave model equations derived by del-Castillo-Negrete \cite{DelCastillo} in an equivalent form by using a class of exact solutions to the Vlasov equation corresponding to the waterbag model \cite{TzenovPRSTAB,TzenovBOOK}. Further, the exact closure of hydrodynamic equations are manipulated following the renormalization group (RG) approach \cite{TzenovBOOK}. As a result, amplitude equations for the slowly varying wave envelopes (SVWE) are derived. Depending on the character of the solutions of the dispersion equation, two cases can be distinguished. In the first case, where all three roots of the dispersion equation are real, the amplitude equations represent a system of three coupled nonlinear equations, describing the process of three-wave interaction and mixing. If the dispersion equation possesses one real and two complex conjugate roots, the amplitude equations take the form of two globally coupled nonlinear equations with complex coefficients. In Section V, the predictions of the SVWE equations are benchmarked to the exact numerical solution of the hydrodynamic equations. Finally, in the last section we draw some conclusions and outlook.

\section{\label{sec:basic}Theoretical Model and Basic Equations}

The basic equations derived by del-Castillo-Negrete \cite{DelCastillo}, which will be the starting point of our subsequent analysis can be written as
\begin{equation}
\partial_t f + {\cal V} \partial_x f + \partial_x \varphi \partial_{\cal V} f = 0, \label{Vlasov}
\end{equation}
\begin{equation}
\varphi = a (t) e^{i x} + a^{\ast} (t) e^{- i x}, \label{Potential}
\end{equation}
\begin{equation}
\sigma {\frac {{\rm d} a} {{\rm d} t}} + i l a = i {\left \langle f e^{- i x} \right \rangle}, \label{PotPoisson}
\end{equation}
where the operator averaging is denoted by
\begin{equation}
{\left \langle \dots \right \rangle} = {\frac {1} {2 \pi}} \int \limits_{- \infty}^{\infty} {\rm d} {\cal V} \int \limits_{0}^{2 \pi} {\rm d} x \dots. \label{Averag}
\end{equation}
The independent time $t$ and spatial $x$ variables, as well as the dependent ones $f {\left( x, {\cal V}; t \right)}$ and $\varphi {\left( x; t \right)}$ entering the above equations have been properly nondimensionalized \cite{DelCastillo}. For example, the dimensionless time $t$ scales as $\omega_e T$, where $\omega_e$ is the electron plasma frequency and $T$ is the real time. Moreover, the velocity ${\cal V}$ in the equations above scales as $V / V_D$, where $V_D$ is a velocity characteristic for the particular problem and can be chosen arbitrarily.

For present purposes, we restrict the analysis to a special class of exact solutions to Eq. (\ref{Vlasov}) corresponding to the waterbag distribution \cite{TzenovPRSTAB,TzenovBOOK}
\begin{equation}
f {\left( x, {\cal V}; t \right)} = {\cal C} {\left[ {\cal H} {\left( {\cal V} - v_{-} {\left( x; t \right)} \right)} - {\cal H} {\left( {\cal V} - v_{+} {\left( x; t \right)} \right)} \right]}, \label{Waterbag}
\end{equation}
where ${\cal H}$ denotes the well-known Heaviside function, ${\cal C}$ is a normalization constant and $0 < x < 2 \pi$ is a normalized spatial variable. It simply means that the phase space density $f {\left( x, {\cal V}; t \right)}$ remains constant within a region confined by the boundary curves $v_{\pm} {\left( x; t \right)}$, which are assumed to be single-valued. The latter distort nonlinearly during the evolution of the system as specified by Eqs. (\ref{Vlasov}) - (\ref{PotPoisson}).

It is convenient to introduce the macroscopic fluid variables
\begin{equation}
\varrho = \int \limits_{- \infty}^{\infty} {\rm d} {\cal V} f {\left( x, {\cal V}; t \right)} = {\cal C} {\left( v_{+} - v_{-} \right)}, \label{Density}
\end{equation}
\begin{equation}
\varrho v = \int \limits_{- \infty}^{\infty} {\rm d} {\cal V} {\cal V} f {\left( x, {\cal V}; t \right)} = {\frac {\cal C} {2}} {\left( v_{+}^2 - v_{-}^2 \right)}, \label{Velocity}
\end{equation}
where $\varrho {\left( x; t \right)}$ and $v {\left( x; t \right)}$ are the density and the current velocity, respectively. It can be shown that the higher moments defining the particle pressure ${\cal P} {\left( x; t \right)}$ and the heat flow ${\cal Q} {\left( x; t \right)}$ can be expressed as
\begin{equation}
{\cal P} = \int \limits_{- \infty}^{\infty} {\rm d} {\cal V} {\left( {\cal V} - v \right)}^2 f {\left( x, {\cal V}; t \right)} = {\frac {{\cal C}} {12}} {\left( v_{+} - v_{-} \right)}^3, \label{Pressure}
\end{equation}
\begin{equation}
{\cal Q} = \int \limits_{- \infty}^{\infty} {\rm d} {\cal V} {\left( {\cal V} - v \right)}^3 f {\left( x, {\cal V}; t \right)} = 0. \label{HeatFlow}
\end{equation}

In particular, the last expression (\ref{HeatFlow}) implies that the waterbag distribution yields an exact closure of the hydrodynamic equations in the form
\begin{equation}
\partial_t \varrho + \partial_x {\left( \varrho v \right)} = 0, \label{Continuity}
\end{equation}
\begin{equation}
\partial_t v + v \partial_x v + v_T^2 \partial_x {\left( \varrho^2 \right)} = \partial_x \varphi, \label{MomBalance}
\end{equation}
where
\begin{equation}
v_T^2 = {\frac {1} {8 {\cal C}^2}}, \label{ThermVelocity}
\end{equation}
is the normalized thermal speed-squared. The macroscopic fluid equations (\ref{Continuity}) and (\ref{MomBalance}) must be supplemented with the equation for the amplitude of the field mode (\ref{PotPoisson}), written in the form
\begin{equation}
\sigma {\frac {{\rm d} a} {{\rm d} t}} + i l a = i {\left \langle \varrho e^{- i x} \right \rangle}, \label{FieldMode}
\end{equation}
where now the operator averaging specified by Eq. (\ref{Averag}) involves integration on the normalized spatial variable $x$ only, and the potential $\varphi$ is given by expression (\ref{Potential}) as before.

We further scale the hydrodynamic and field variables according to the expressions
\begin{equation}
\varrho = \varrho_0 + \epsilon R, \quad v = v_0 + \epsilon V, \quad a = \epsilon \alpha, \quad \varphi = \epsilon \Phi, \label{Scaling}
\end{equation}
where $\epsilon$ is a formal small parameter, and will be set equal to one at the end of the calculations. Furthermore, $\varrho_0 = const$ and $v_0 = const$ represent the stationary solution of Eqs. (\ref{Continuity}) and (\ref{MomBalance}), provided the stationary field amplitude $a_0 = 0$. The basic macroscopic fluid and electrostatic field equations can be rewritten
as
\begin{equation}
\partial_t R + \varrho_0 \partial_x V + v_0 \partial_x R = - \epsilon \partial_x {\left( R V \right)}, \label{ContinScale}
\end{equation}
\begin{equation}
\partial_t V + v_0 \partial_x V + 2 \varrho_0 v_T^2 \partial_x R - \partial_x \Phi = - \epsilon \partial_x {\left( {\frac {V^2} {2}} + v_T^2 R^2 \right)}, \label{MomBalScale}
\end{equation}
\begin{equation}
\sigma {\frac {{\rm d} \alpha} {{\rm d} t}} + i l \alpha = i {\left \langle R e^{- i x} \right \rangle}, \label{FieldMoScale}
\end{equation}
\begin{equation}
\Phi = \alpha (t) e^{i x} + \alpha^{\ast} (t) e^{- i x}. \label{PotenScale}
\end{equation}
To simplify the above system of equations, we perform a Galilean transformation specified by
\begin{equation}
z = x - v_0 t, \qquad \Psi (t) = \alpha (t) e^{i v_0 t}, \label{Galilean}
\end{equation}
and cast our basic system of equations in the form
\begin{equation}
\partial_t R + \varrho_0 \partial_z V = - \epsilon \partial_z {\left( R V \right)}, \label{ContinuG}
\end{equation}
\begin{equation}
\partial_t V + 2 \varrho_0 v_T^2 \partial_z R - \partial_z \Phi = - \epsilon \partial_z {\left( {\frac {V^2} {2}} + v_T^2 R^2 \right)}, \label{MomBalaG}
\end{equation}
\begin{equation}
\sigma {\frac {{\rm d} \Psi} {{\rm d} t}} + i {\cal L} \Psi = i {\left \langle R e^{- i z} \right \rangle}, \label{FieldModG}
\end{equation}
\begin{equation}
\Phi = \Psi (t) e^{i z} + \Psi^{\ast} (t) e^{- i z}, \qquad {\cal L} = l - \sigma v_0. \label{PotentG}
\end{equation}
Eliminating $V$ from the left-hand-sides of Eqs. (\ref{ContinuG}) and (\ref{MomBalaG}), we arrive at the basic system
\begin{equation}
\partial_t^2 R - \lambda^2 \partial_z^2 R - \varrho_0 \Phi = - \epsilon \partial_t \partial_z {\left( R V \right)} + \epsilon \varrho_0 \partial_z^2 {\left( {\frac {V^2} {2}} + v_T^2 R^2 \right)}, \label{BasicEq1}
\end{equation}
\begin{equation}
\sigma \partial_t \Psi + i {\cal L} \Psi = i {\left \langle R e^{- i z} \right \rangle}, \quad \qquad \lambda^2 = 2 \varrho_0^2 v_T^2, \label{BasicEq2}
\end{equation}
for the subsequent analysis using the RG approach.

\section{\label{sec:reduction}Renormalization Group Reduction of the Macroscopic Equations}

Following the standard procedure of the RG method \cite{TzenovBOOK}, we represent ${\widehat{\cal G}} {\left( z; t \right)}$ as a perturbation expansion
\begin{equation}
{\widehat{\cal G}} {\left( z; t \right)} = \sum \limits_{n=0}^{\infty} \epsilon^n {\widehat{\cal G}}_n {\left( z; t \right)}, \label{StandardPert}
\end{equation}
where ${\widehat{\cal G}} = {\left( R, V, \Phi \right)}$ represents all hydrodynamic and field variables. To zero order, the perturbation equations (\ref{BasicEq1}) and (\ref{BasicEq2}) can be written as
\begin{equation}
\partial_t^2 R_0 - \lambda^2 \partial_z^2 R_0 = \varrho_0 {\left( \Psi_0 e^{iz} + \Psi_0^{\ast} e^{-iz} \right)}, \label{BasicEq1Zero}
\end{equation}
\begin{equation}
\sigma \partial_t \Psi_0 + i {\cal L} \Psi_0 = i {\left \langle R_0 e^{- i z} \right \rangle}. \label{BasicEq2Zero}
\end{equation}
Since the equation for $R_0$ contains only the first harmonic with regard to the spatial variable $z$ [proportional to $\exp (\pm i z)$], its solution is sought in the form
\begin{equation}
R_0 {\left( z; t \right)} = {\cal F} {\left( t \right)} e^{iz} + {\cal F}^{\ast} {\left( t \right)} e^{-iz}, \label{SolSoughtR0}
\end{equation}
where the function ${\cal F}$ satisfies the equation
\begin{equation}
{\left( \partial_t^2 + \lambda^2 \right)} {\left( \sigma \partial_t + i {\cal L} \right)} {\cal F} - i \varrho_0 {\cal F} = 0. \label{EquationF}
\end{equation}
Taking the latter into account, we can write
\begin{equation}
R_0 {\left( z; t \right)} = \sum \limits_m {\cal A}_m e^{i \omega_m t + iz} + \sum \limits_m {\cal A}_m^{\ast} e^{- i \omega_m^{\ast} t - iz}, \label{GenSolR0}
\end{equation}
where the sum spans over all roots of the characteristic equation
\begin{equation}
{\left( \omega^2 - \lambda^2 \right)} {\left( \sigma \omega + {\cal L} \right)} + \varrho_0 = 0. \label{CharEquat}
\end{equation}
In general, the characteristic equation has three roots. However, as it will become clear from the subsequent exposition, in a number of cases depending on the particular values of physical parameters one of the roots is real, while the other two are complex conjugate. Note also that the arbitrary (to this end) amplitudes ${\cal A}_m$ are constants with respect to the spatial $z$ and time $t$ variables. Using Eqs. (\ref{MomBalaG}) and (\ref{BasicEq2Zero}), we find
\begin{equation}
V_0 = - {\frac {1} {\varrho_0}} \sum \limits_m \omega_m {\cal A}_m e^{i \omega_m t + iz} - {\frac {1} {\varrho_0}} \sum \limits_m \omega_m^{\ast} {\cal A}_m^{\ast} e^{- i \omega_m^{\ast} t - iz}, \label{GenSolV0}
\end{equation}
\begin{equation}
\Psi_0 = {\frac {1} {\varrho_0}} \sum \limits_m {\left( \lambda^2 - \omega_m^2 \right)} {\cal A}_m e^{i \omega_m t}. \label{GenSolPsi0}
\end{equation}
It is worthwhile mentioning that in general, the solution of Eq. (\ref{BasicEq1Zero}) would contain all other harmonics starting with the second one [proportional to $\exp (\pm i n z)$, where $n$ is integer]. They are "sound" waves, propagating at speed $\lambda$ and are decoupled from the electric field amplitude $\Psi$. Their amplitudes are vanishing in the zero order approximation. This assertion will be verified numerically and will be discussed in more detail in Section V.

In first order the basic equations (\ref{BasicEq1}) and (\ref{BasicEq2}) can be expressed as
\begin{equation}
\partial_t^2 R_1 - \lambda^2 \partial_z^2 R_1 - \varrho_0 \Phi_1 = \nonumber
\end{equation}
\begin{equation}
- {\frac {1} {\varrho_0}} \sum \limits_{m,n} G_{mn} {\cal A}_m {\cal A}_n e^{i {\left( \omega_m + \omega_n \right)} t + 2iz} + c.c., \label{BasicEq1First}
\end{equation}
\begin{equation}
\sigma \partial_t \Psi_1 + i {\cal L} \Psi_1 = i {\left \langle R_1 e^{- i z} \right \rangle}. \label{BasicEq2First}
\end{equation}
Here and afterwards, the symbol "$c.c$" stands for the complementary complex conjugate counterpart, while $G_{mn}$ is a symmetric matrix given by the expression
\begin{equation}
G_{mn} = {\left( \omega_m + \omega_n \right)}^2 + 2 {\left( \omega_m \omega_n + \lambda^2 \right)}. \label{SymMatG}
\end{equation}
It can be verified in a straightforward manner that the general solution of the first order equations is
\begin{equation}
R_1 = \sum \limits_{\nu \neq \pm 1} {\cal B}_{\nu}^{(+)} e^{i \nu (z + \lambda t)} + \sum \limits_{\nu \neq \pm 1} {\cal B}_{\nu}^{(-)} e^{i \nu (z - \lambda t)} \nonumber
\end{equation}
\begin{equation}
+ {\frac {1} {\varrho_0}} \sum \limits_{m,n} F_{mn} {\cal A}_m {\cal A}_n e^{i {\left( \omega_m + \omega_n \right)} t + 2iz} + c.c., \label{GenSolEq1First}
\end{equation}
where
\begin{equation}
F_{mn} = {\frac {{\left( \omega_m + \omega_n \right)}^2 + 2 {\left( \omega_m \omega_n + \lambda^2 \right)}} {{\left( \omega_m + \omega_n \right)}^2 - 4 \lambda^2}}, \label{SymMatF}
\end{equation}
and ${\cal B}_{\nu}^{(\pm)}$ are the arbitrary constant amplitudes of the solution to the homogeneous part of Eq. (\ref{BasicEq1First}). The summation over $\nu$ runs from $- \infty$ to $\infty$ excluding $\nu = \pm 1$.  The first harmonic [proportional to $\exp(\pm iz)$] describes the coupling between the fluid dynamic and the electric fields and has been taken into account in zero order. For the first order current velocity $V_1$ we obtain
\begin{equation}
V_1 = - {\frac {\lambda} {\varrho_0}} {\left[ \sum \limits_{\nu \neq \pm 1} {\cal B}_{\nu}^{(+)} e^{i \nu (z + \lambda t)} - \sum \limits_{\nu \neq \pm 1} {\cal B}_{\nu}^{(-)} e^{i \nu (z - \lambda t)} \right]} \nonumber
\end{equation}
\begin{equation}
- {\frac {1} {\varrho_0^2}} \sum \limits_{m,n} {\cal V}_{mn} {\cal A}_m {\cal A}_n e^{i {\left( \omega_m + \omega_n \right)} t + 2iz} + c.c., \label{GenSolEqVFirst}
\end{equation}
where
\begin{equation}
{\cal V}_{mn} = {\frac {\omega_m \omega_n + \lambda^2 + 2 \lambda^2 F_{mn}} {\omega_m + \omega_n}}. \label{SymMatV}
\end{equation}
Note that in first order the electric field vanishes $\Psi_1 = 0$.

The final step in our perturbative procedure is to obtain the secular second-order solution. Retaining terms giving rise to secular contributions in the second-order solution, we can express the constitutive equations
as follows
\begin{equation}
\partial_t^2 R_2 - \lambda^2 \partial_z^2 R_2 - \varrho_0 \Phi_2 = \nonumber
\end{equation}
\begin{equation}
- {\frac {1} {\varrho_0^2}} \sum \limits_{m,n,l} \Delta_{nlm} {\cal A}_m^{\ast} {\cal A}_n {\cal A}_l  e^{i {\left( \omega_n + \omega_l - \omega_m^{\ast} \right)} t + iz} + c.c., \label{BasicEq1Second}
\end{equation}
\begin{equation}
\sigma \partial_t \Psi_2 + i {\cal L} \Psi_2 = i {\left \langle R_2 e^{- i z} \right \rangle}, \label{BasicEq2Second}
\end{equation}
where the coupling coefficients $\Delta_{mnl}$ is given by the expression
\begin{equation}
\Delta_{nlm} = \omega_n \omega_l + \lambda^2 + 3 \lambda^2 F_{nl} \nonumber
\end{equation}
\begin{equation}
+ {\left( \omega_n + \omega_l - \omega_m^{\ast} \right)} \omega_m^{\ast} F_{nl}, \label{SymMatGamma}
\end{equation}
Taking into account terms driving self- and cross-modulation, it is straightforward to verify that the secular second-order solution can be expressed as
\begin{equation}
R_2 = {\frac {it} {\varrho_0^2 \lambda^2}} \sum \limits_{m,n} V_{gm} \Upsilon_{mn} {\cal A}_m {\left| {\widetilde{\cal A}}_n \right|}^2 e^{i \omega_m t + iz} + c.c., \label{GenSolEq1Second}
\end{equation}
where
\begin{equation}
\Upsilon_{mn} = \Gamma_{mn} {\frac {\sigma {\left( \omega_m + 2i \gamma_n \right)} + {\cal L}} {\sigma \omega_m + {\cal L}}} \qquad \Gamma_{mn} = \Delta_{mnn}, \label{BasicCoupMatrix}
\end{equation}
and
\begin{equation}
\gamma_m = {\rm Im} {\left( \omega_m \right)}, \qquad {\widetilde{\cal A}}_m = {\cal A}_m e^{- \gamma_m t}. \label{Additional}
\end{equation}
Moreover, the quantity $V_{gm}$ is the group velocity defined as follows. Let us introduce the dispersion function \cite{TzenovBOOK}
\begin{equation}
{\cal D} {\left( k, \omega \right)} = {\left( \lambda^2 k^2 - \omega^2 \right)} {\left( \sigma \omega + {\cal L} \right)} - \varrho_0, \label{DisperFunc}
\end{equation}
corresponding to a general solution of Eq. (\ref{BasicEq1}) proportional to $\exp{\left( i \omega t + ikz \right)}$, where $k$ is the wave number. Obviously, the dispersion equation for $k =1$ that is, ${\cal D} {\left( 1, \omega \right)} = 0$ reduces to the characteristic equation (\ref{CharEquat}). The group velocity is given by the expression \cite{TzenovBOOK}
\begin{equation}
V_{gm} = {\left. {\frac {{\rm d} \omega_m} {{\rm d} k}} \right|}_{k=1} = - {\left. {\frac {\partial {\cal D}} {\partial k}} {\left( {\frac {\partial {\cal D}} {\partial \omega_m}} \right)}^{-1} \right|}_{k=1}, \label{GroupVelDef}
\end{equation}
or in an explicit form
\begin{equation}
V_{gm} = {\frac {2 \varrho_0 \lambda^2} {{\left( \omega_m^2 - \lambda^2 \right)} {\left( \sigma \lambda^2 - 3 \sigma \omega_m^2 - 2 {\cal L} \omega_m \right)}}}. \label{GroupVel}
\end{equation}
The last step is to collect all terms corresponding to increasing orders, which contribute to $R {\left( z; t \right)}$ and perform a resummation such as to absorb secular terms (proportional to various powers of the time variable $t$) present in $R_2$. Since this approach is standard \cite{TzenovBOOK}, we omit details here.

\section{\label{sec:rgequation}The Renormalization Group Equation}

Following the procedure \cite{TzenovBOOK} of the RG method, we finally obtain the desired RG equation
\begin{equation}
i \partial_t {\widetilde{\cal A}}_m  = - {\frac {V_{gm}} {\varrho_0^2 \lambda^2}} \sum \limits_n \Upsilon_{mn} {\widetilde{\cal A}}_m {\left| {\widetilde{\cal A}}_n \right|}^2, \label{RGEquation}
\end{equation}
where now ${\widetilde{\cal A}}_m$ denotes the renormalized complex amplitude of the type (\ref{Additional}). In terms of the renormalized wave envelopes up to first order in the formal expansion parameter, the macroscopic density $\varrho {\left( z; t \right)}$ can be expressed as
\begin{equation}
\varrho {\left( z; t \right)} = \varrho_0 + \sum \limits_m {\widetilde{\cal A}}_m {\left( t \right)} e^{i {\rm Re} {\left( \omega_m \right)} t + iz} \nonumber
\end{equation}
\begin{equation}
+ \sum \limits_m {\widetilde{\cal A}}_m^{\ast} {\left( t \right)} e^{- i {\rm Re} {\left( \omega_m \right)} t - iz}. \label{RenDensity}
\end{equation}
Similar expressions hold for the current velocity $V$ [compare with Eq. (\ref{GenSolV0})] and for the electrostatic potential $\Psi$ [compare with Eq. (\ref{GenSolPsi0})].

It was mentioned in the previous section that since the characteristic equation (\ref{CharEquat}) is an algebraic equation of third order, it possesses three roots in general. Thus, we can distinguish the following two cases of physical interest.

All three roots $\omega_1$, $\omega_2$ and $\omega_3$ are real. As it will be shown in the next Section, at certain value of the parameter $v_0$ they satisfy the resonance condition
\begin{equation}
\omega_1 + \omega_2 = 2 \omega_3 + \delta. \label{Resonance}
\end{equation}
Apart from the self and cross modulation terms in Eq. (\ref{RGEquation}), additional terms describing wave mixing will be present. Therefore, in the most general case the renormalization group equation can be written as
\begin{equation}
i \partial_t {\cal A}_1 = - {\frac {V_{g1}} {\varrho_0^2 \lambda^2}} {\left( \Gamma_{11}  {\left| {\cal A}_1 \right|}^2 + \Gamma_{12}  {\left| {\cal A}_2 \right|}^2 + \Gamma_{13}  {\left| {\cal A}_3 \right|}^2 \right)} {\cal A}_1 \nonumber
\end{equation}
\begin{equation}
= - {\frac {V_{g1}} {\varrho_0^2 \lambda^2}} {\left[ \omega_3^2 + \lambda^2 + {\left( 3 \lambda^2 + \omega_1 \omega_2 \right)} F_{33} \right]} {\cal A}_2^{\ast} {\cal A}_3^2 e^{- i \delta t}, \label{RGEquationReal1}
\end{equation}
\begin{equation}
i \partial_t {\cal A}_2 = - {\frac {V_{g2}} {\varrho_0^2 \lambda^2}} {\left( \Gamma_{21}  {\left| {\cal A}_1 \right|}^2 + \Gamma_{22}  {\left| {\cal A}_2 \right|}^2 + \Gamma_{23}  {\left| {\cal A}_3 \right|}^2 \right)} {\cal A}_2 \nonumber
\end{equation}
\begin{equation}
= - {\frac {V_{g2}} {\varrho_0^2 \lambda^2}} {\left[ \omega_3^2 + \lambda^2 + {\left( 3 \lambda^2 + \omega_1 \omega_2 \right)} F_{33} \right]} {\cal A}_1^{\ast} {\cal A}_3^2 e^{- i \delta t}, \label{RGEquationReal2}
\end{equation}
\begin{equation}
i \partial_t {\cal A}_3 = - {\frac {V_{g3}} {\varrho_0^2 \lambda^2}} {\left( \Gamma_{31}  {\left| {\cal A}_1 \right|}^2 + \Gamma_{32}  {\left| {\cal A}_2 \right|}^2 + \Gamma_{33}  {\left| {\cal A}_3 \right|}^2 \right)} {\cal A}_3 \nonumber
\end{equation}
\begin{equation}
= - {\frac {V_{g3}} {\varrho_0^2 \lambda^2}} {\left( \omega_1 \omega_2 + \lambda^2 + 3 \lambda^2 F_{12} + \omega_3^2 F_{13} \right)} {\cal A}_1 {\cal A}_2 {\cal A}_3^{\ast} e^{i \delta t}. \label{RGEquationReal3}
\end{equation}
The terms proportional to $\Gamma_{nn}$ describe the effects of phase self-modulation, whereas those proportional to $\Gamma{mn}$ (with $m \neq n$) are responsible for phase cross-modulation. Both result in an intensity-dependent, nonlinear frequency shift of the eigenfrequencies $\omega_m$, an effect similar to the nonlinear wave-number shift in nonlinear optics \cite{Agrawal}. In contrast, the last terms in each of the equations above describe four-wave mixing effects, responsible for energy exchange between the three modes. These effects play a role only when the phase mismatch parameter $\delta$ is close to zero.

One real $\omega_1$ root and two $\omega_2$ and $\omega_2^{\ast}$ complex conjugate roots. In this case the
renormalization group equation (\ref{RGEquation}) takes the form
\begin{equation}
i \partial_t {\cal A}_1 = - {\frac {V_{g1}} {\varrho_0^2 \lambda^2}} {\left[ \Gamma_{11} {\left| {\cal A}_1 \right|}^2 + \Gamma_{12} {\left( 1 + 2 i \gamma_{12} \right)} {\left| {\widetilde{\cal A}}_2 \right|}^2 \right]} {\cal A}_1, \label{RGEquationGL1}
\end{equation}
\begin{equation}
i \partial_t {\widetilde{\cal A}}_2 = - {\frac {V_{g2}} {\varrho_0^2 \lambda^2}} {\left[ \Gamma_{21} {\left| {\cal A}_1 \right|}^2 + \Gamma_{22} {\left( 1 + 2 i \gamma_{22} \right)} {\left| {\widetilde{\cal A}}_2 \right|}^2 \right]} {\widetilde{\cal A}}_2, \label{RGEquationGL2}
\end{equation}
where the quantities $\gamma_{12}$ and $\gamma_{22}$ are given by the expressions
\begin{equation}
\gamma_{12} = {\frac {\sigma \gamma_2} {\sigma \omega_1 + {\cal L}}}, \quad \qquad \gamma_{22} = {\frac {\sigma \gamma_2} {\sigma \omega_2 + {\cal L}}}. \label{Constant1}
\end{equation}

An important comment is now in order. The equations for the slowly varying wave amplitudes describing the coupling between the hydrodynamic and electric fields were obtained in third order in the formal small parameter. Note that up to this order the "sound" waves with amplitudes ${\cal B}_{\nu}^{(\pm)}$ remain uncoupled with the basic modes with amplitudes ${\cal A}_m$. Such coupling (similar to the wave mixing of the three eigenmodes) may take place provided a resonance condition of the form $\omega_m + \omega_n = \pm 2 \lambda$ is fulfilled, where $\omega_m$ and $\omega_n$ are any two roots of the characteristic equation (\ref{CharEquat}). Note also that the amplitudes of the "sound" waves remain undefined to third order in the formal expansion parameter. To obtain the corresponding amplitude equations for these modes, one has to go beyond third order, which is out of the scope of the present paper.

\section{\label{sec:simulation}Numerical Simulations}

In this Section we present a comparison between the numerical solution of the system of hydrodynamic equations (\ref{Continuity}), (\ref{MomBalance}) supplemented with the equation for the single wave amplitude (\ref{FieldMode}), and the RG solution obtained in the previous section. Since the nondimensionalization of the initial phase space variables allows an arbitrary velocity scale, it is convenient to define the latter such that
\begin{equation}
v_T^2 = {\frac {3} {2}}. \label{Scalin}
\end{equation}

In order to assess the accuracy of the model equations (\ref{RGEquationReal1}) - (\ref{RGEquationReal3}) the predictions of the latter are benchmarked to the numerical solution of the original system of Eqs. (\ref{Continuity}), (\ref{MomBalance}) and (\ref{FieldMode}). To achieve this an initial condition of the form $\varrho {\left( x; t=0 \right)} = 1 + \epsilon \cos(x)$ with $\epsilon$ in the range $0.001 \dots 0.01$ and $v {\left( x; t=0 \right)} = v_0 = const(x)$ has been used. Equations (\ref{GenSolR0}), (\ref{GenSolV0}) and (\ref{GenSolPsi0}) relate the initial distributions of $\varrho {\left( x; t=0 \right)}$, $v {\left( x; t=0 \right)}$ and $\Psi {\left( t=0 \right)}$ to the three slowly varying amplitudes ${\cal A}_1{\left( t=0 \right)}$, ${\cal A}_2{\left( t=0 \right)}$ and ${\cal A}_3{\left( t=0 \right)}$. Varying $\epsilon$ in the range $10^{-3} \dots 10^{-2}$ corresponds to a transition from a linear to a weakly nonlinear system dynamics. Periodic boundary conditions have been implemented.

\begin{figure}
\begin{center}
\includegraphics[width=8.0cm]{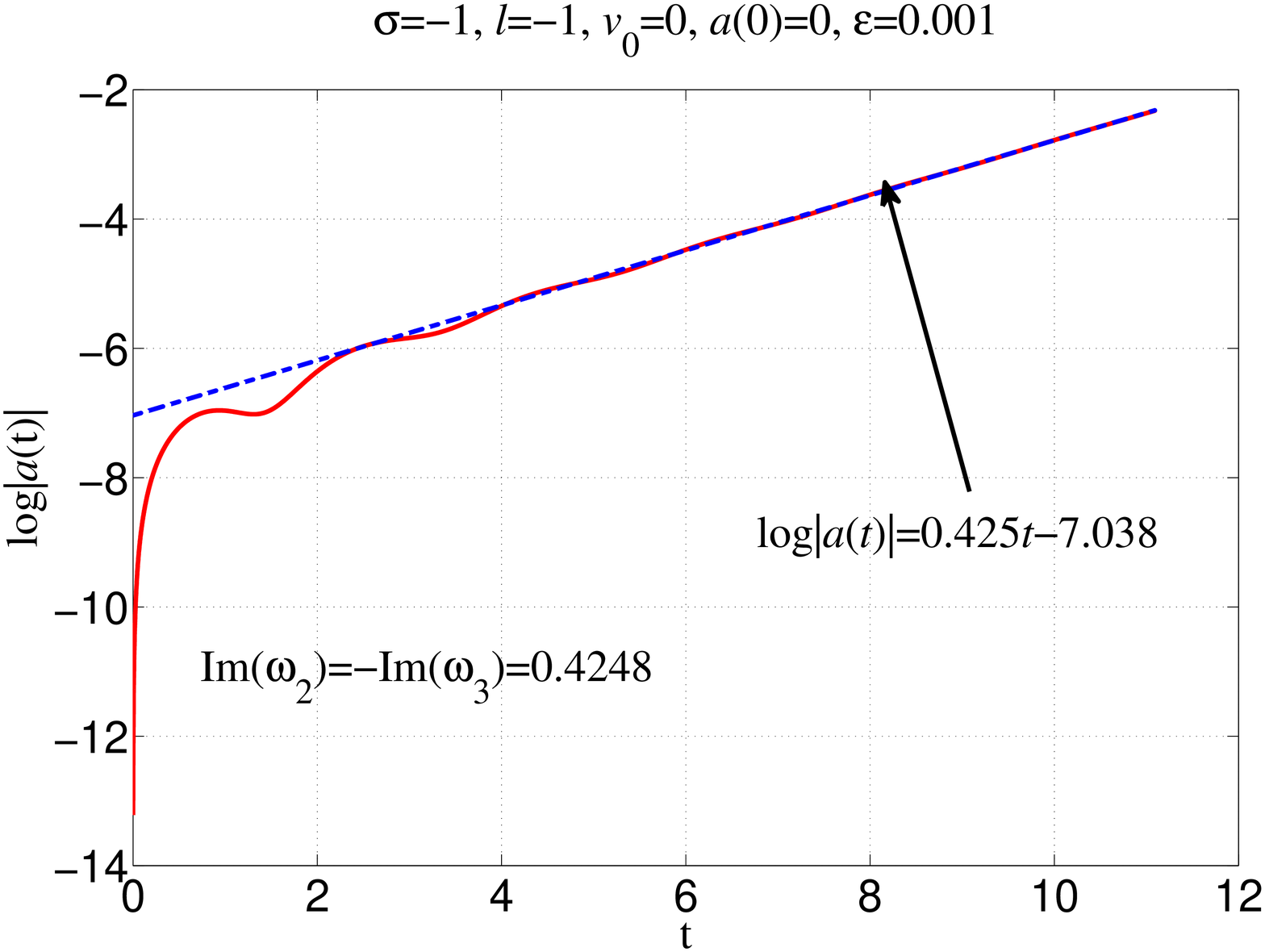}
\caption{\label{fig:epsart1} Continuous line: time-dependence of the amplitude $a(t)$ at $l = -1$ and $\sigma = -1$ (one real and two complex roots of the characteristic equation) obtained by numerical integration of Eqs. (\ref{Continuity}), (\ref{MomBalance}) and (\ref{FieldMode}). The asymptotic behavior of $a(t)$ is given by the dashed line. Its slope is the absolute value of the imaginary part of the complex roots.}
\end{center}
\end{figure}

\begin{figure}
\begin{center}
\includegraphics[width=8.0cm]{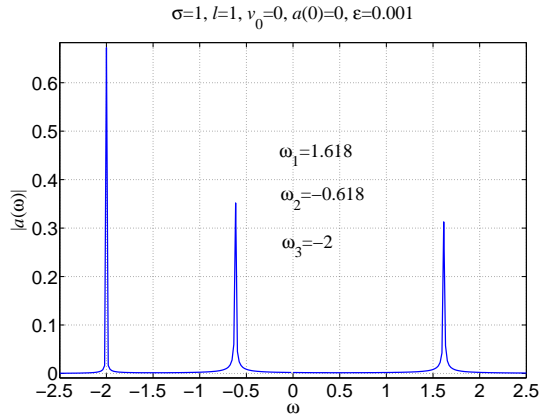}
\caption{\label{fig:epsart2} Spectrum $a(\omega)$ of the amplitude $a(t)$ in the case of $l = 1$ and $\sigma = 1$ obtained from the numerical solution of Eqs. (\ref{Continuity}), (\ref{MomBalance}) and (\ref{FieldMode}). The three peaks correspond to the three real roots of the characteristic equation. Note the absence of "sound" waves at frequencies $\omega = \pm {\sqrt{3}}$.}
\end{center}
\end{figure}

The results shown in Figures \ref{fig:epsart1} and \ref{fig:epsart2} pertain to the linear dynamics regime. It can be shown that in the case, where $l = -1$ and $\sigma = -1$ (Fig. 1) the characteristic equation (\ref{CharEquat}) has one real and two complex conjugate roots. From Eq. (\ref{GenSolPsi0}) it is clear that the mode corresponding to the root with negative imaginary part will become dominant for sufficiently late times $t$. This observation is in full agreement with Fig. 2. In addition, the slope of the asymptotic line matches the imaginary part of the complex roots with an accuracy of the order of $10^{-4}$. In contrast, when $l = 1$ and $\sigma = 1$ the characteristic equation possesses three real roots and these correspond to the three eigenfrequencies of the system $\omega_1$, $\omega_2$ and $\omega_3$. By Fourier-analyzing the numerical data for the amplitude of the electrostatic potential $a(t)$, the values of these frequencies can be obtained and compared to the roots of the characteristic equation. As Fig. \ref{fig:epsart2} shows the two results are in excellent agreement.

Increasing $\epsilon$ from $10^{-3}$ to $10^{-2}$ effectively switches on the nonlinearity. It can be shown that the phase-matching (resonance) condition (\ref{Resonance}) is satisfied at $v_{0} = v_{pm} \approx 0.49$. In order to assess the efficiency of the four-wave mixing (FWM) process, Eqs. (\ref{RGEquationReal1}) - (\ref{RGEquationReal3}) have been solved for several values of $v_{0}$ in the vicinity of $v_{pm}$. It was found that significant energy exchange between the modes occurs on time scales of the order of $10^{4}$. Solving the original system of hydrodynamic equations (\ref{Continuity}), (\ref{MomBalance}) and (\ref{FieldMode}) on such a timescale is not practical and given the small efficiency of the FWM process it can be neglected.

\begin{figure}
\begin{center}
\includegraphics[width=8.0cm]{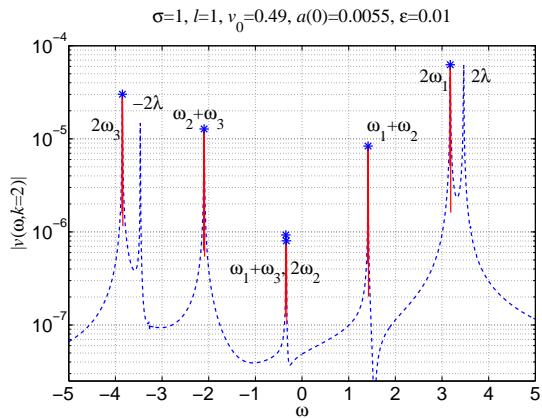}
\caption{\label{fig:epsart3} Fourier transform $v(\omega, k = 2)$ ($k$ is the harmonic number) of the current velocity $v(z; t)$. The dashed line is the spectrum obtained by a fast Fourier transform. In addition "slow" Fourier transform is performed in the vicinity of the peaks in order to improve the frequency resolution and the continuous line is the result. The "*"-symbols represent the analytical result, given by Eq. (\ref{GenSolEqVFirst}). The values of the eigenfrequencies are $\omega_{1} = 1.589$, $\omega_{2} = -0.173$ and $\omega_{3} = -1.925$. The two components at frequencies $-2 \lambda$ and $2 \lambda$ are "sound" waves (harmonic number k = 2),  propagating at speed $\lambda$. }
\end{center}
\end{figure}

\begin{figure}
\begin{center}
\includegraphics[width=8.0cm]{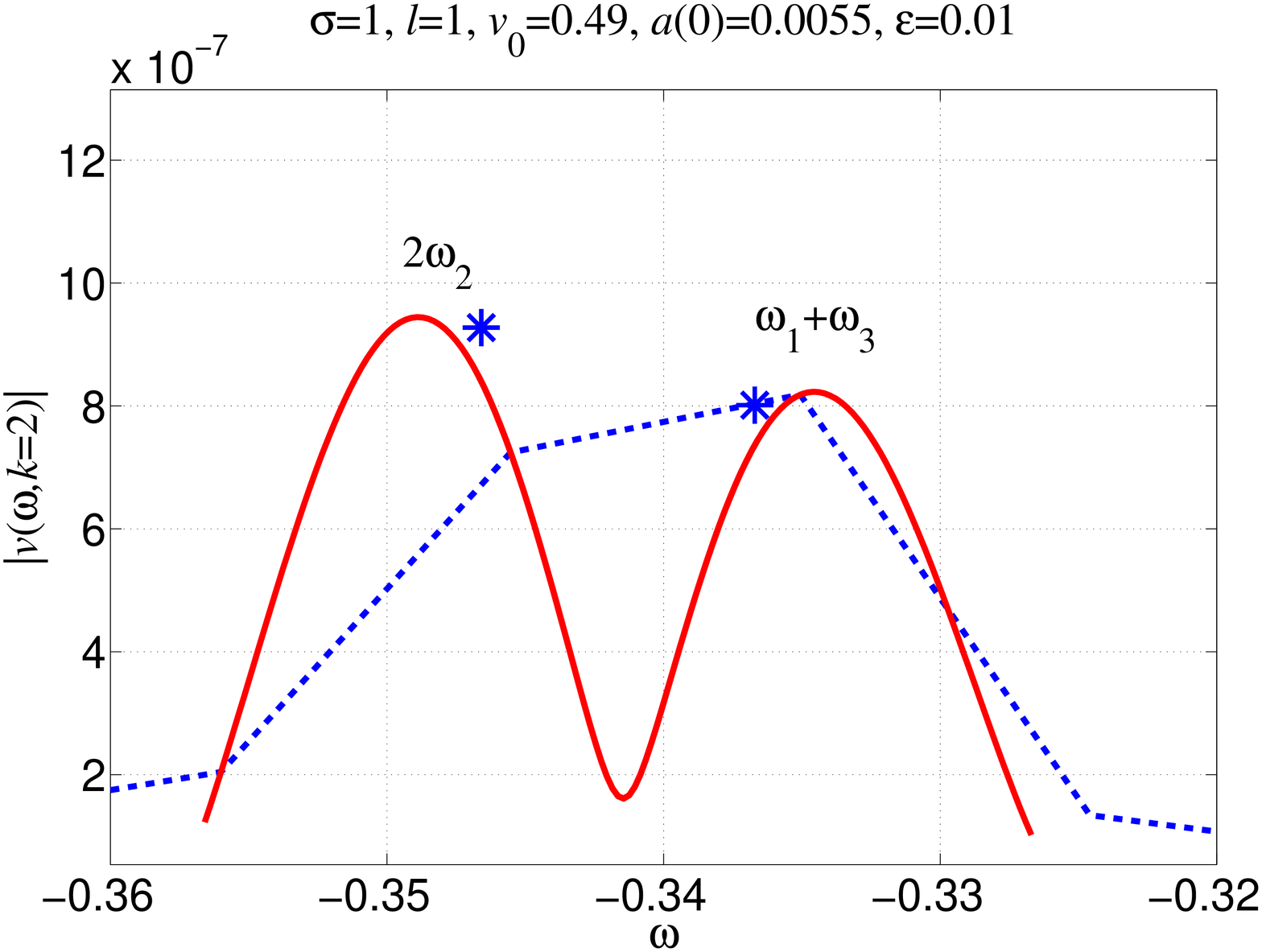}
\caption{\label{fig:epsart4} The same as in Fig. \ref{fig:epsart3} but the frequency region near $2\omega_{2}$ and $\omega_{1}+\omega_{3}$ is shown in more detail. }
\end{center}
\end{figure}

Figures \ref{fig:epsart3} and \ref{fig:epsart4} compare the numerical data for the current velocity $v(z;t)$ with the analytical result, given by Eq. (\ref{GenSolEqVFirst}). Firstly, the quantity $v(\omega, k=2)$ has been computed by fast-Fourier transforming the quantity ${\left \langle v(z; t) \exp(-2iz) \right \rangle}$ and this results in locating the peaks of the spectrum (dashed line in Figs. \ref{fig:epsart3} and \ref{fig:epsart4}). In order to increase the accuracy "slow" Fourier transform has been performed in the vicinity of the peaks (continuous lines). Finally, the "*"-symbols represent the amplitudes of the components at frequencies $\omega_{m} + \omega_{n}$ computed from Eq. (\ref{GenSolEqVFirst}). As can be seen, the analytical and numerical results are in very good agreement. Note, that besides the six frequency components at frequencies $\omega_{m} + \omega_{n}$ there are two additional peaks at frequencies $\omega = \pm 2 \lambda$. They represent "sound" waves, solutions of the Eqs. (\ref{Continuity}), (\ref{MomBalance}) and (\ref{FieldMode}) with $\varphi = 0$ and harmonic number $k = 2$ propagating at the sound velocity $\lambda$. Since these waves are not in resonance with any of the six components at frequencies $\omega_{m} + \omega_{n}$ ($\pm 2 \lambda \neq \omega_{m} + \omega_{n}$) and their amplitudes remain much smaller than ${\cal A}_1$, ${\cal A}_2$ and ${\cal A}_3$, their effect on the evolution of the latter can be neglected.

\begin{figure}
\begin{center}
\includegraphics[width=8.0cm]{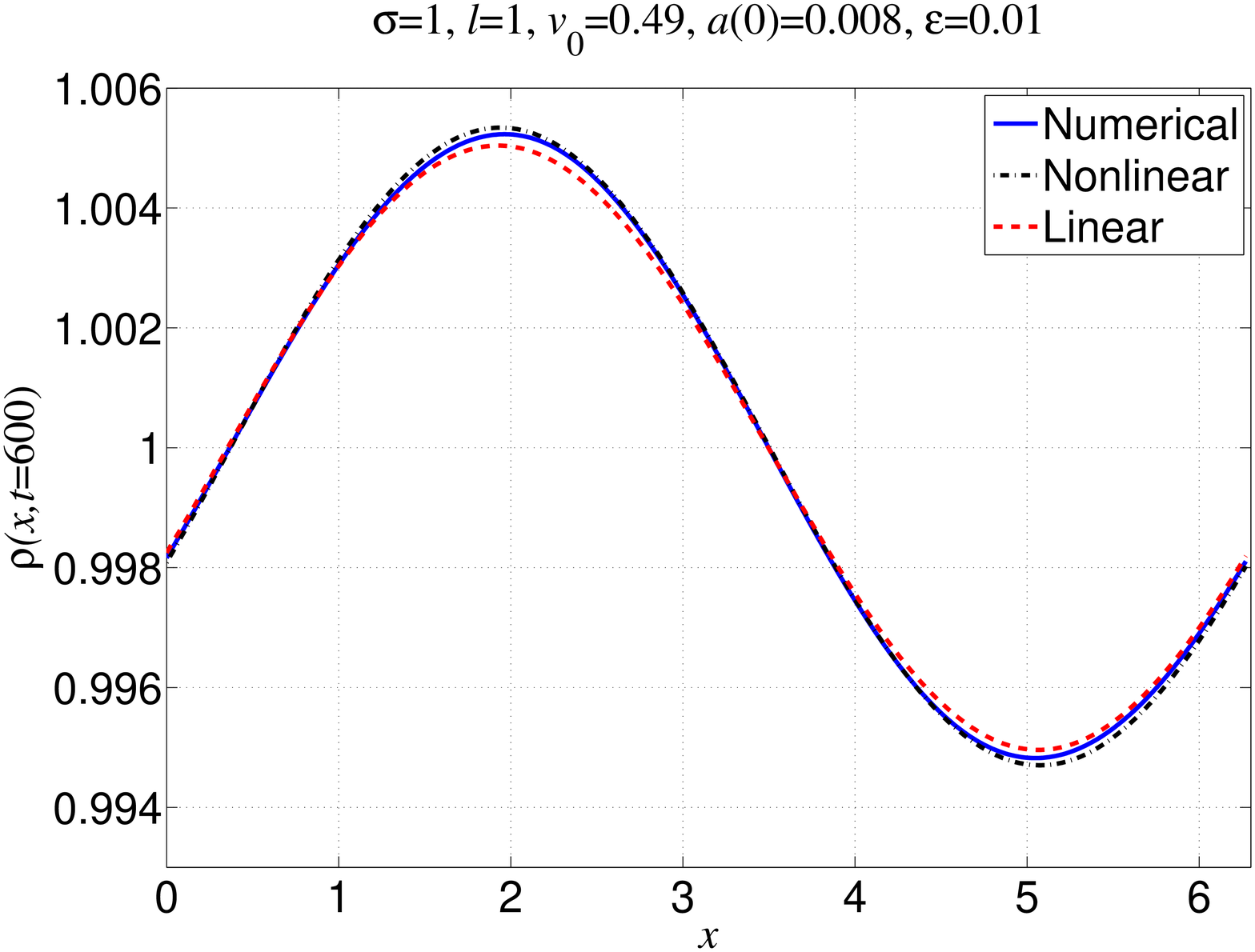}
\caption{\label{fig:epsart5} "Numerical": density distribution $\varrho(x;t=600)$ obtained from the numerical solution of Eqs. (\ref{Continuity}), (\ref{MomBalance}) and (\ref{FieldMode}); "Linear": Eq. (\ref{GenSolR0}) with ${\cal A}_m=const$. The curve marked "Nonlinear" is obtained by taking into account the nonlinear frequency shift of ${\cal A}_m$, Eqs. (\ref{RGEquationReal1}) - (\ref{RGEquationReal3}). }
\end{center}
\end{figure}

Figure \ref{fig:epsart5} compares the analytical and numerical results for the density $\varrho(x; t)$. The curve labeled "Linear" is obtained from Eq. (\ref{GenSolR0}) with ${\cal A}_m = const$. The curve with label "Nonlinear" in addition takes into account the "slow" variation of ${\cal A}_m$, resulting from the nonlinear self- and cross-modulation effects described by Eqs. (\ref{RGEquationReal1}) - (\ref{RGEquationReal3}). As can be seen taking into account the nonlinear effect improves the accuracy of the analytical solution. This has been verified by calculating the square difference between the numerical and the "linear" and "nonlinear" analytical results, respectively.

\section{\label{sec:remarks}Concluding Remarks}

Using a class of waterbag phase space density distributions, which is an exact solutions to the Vlasov equation, we have cast the single-wave model equations to a fluid dynamic form with nonzero thermal velocity. Since the continuity and momentum balance equations comprise an exact hydrodynamic closure, the hydrodynamic representation is fully equivalent to the original system comprising the Vlasov equation and the evolution equation for the single electrostatic field mode.

Based on the renormalization group method, a system of coupled nonlinear equations for the slowly varying amplitudes of interacting plasma density waves has been derived. Depending on the solution of the characteristic equation, the system mentioned above takes the form of either three coupled nonlinear equations describing the process of three-wave interaction in the case, where all three roots are real, or two coupled equations with complex coefficients if the characteristic equation possesses one real and two complex conjugate roots.

In linear dynamic mode (and three real roots of the characteristic equation), the values of the three eigenfrequencies $\omega_1$, $\omega_2$ and $\omega_3$ computed numerically have been found in excellent agreement with the solutions of the characteristic equation (\ref{CharEquat}). Similarly, in the case of a single real root, the asymptotic behavior of the electrostatic potential is in full accordance with the value of imaginary part of the complex roots. In the weakly nonlinear regime, the spectrum of the nonlinear correction to the current velocity closely matches the analytical solution. Finally, direct comparison between the analytical expressions for the density $\varrho (x; t)$, with and without nonlinear effects, and its numerical solution via numerical integration of the hydrodynamic equations shows that taking the nonlinear effects into account improves the agreement between the analytical and numerical results.

It is worthwhile to mention that similar equations have been obtained earlier by Janssen and Rasmussen \cite{Janssen}. Although in a different physical context, amplitude equations similar to the ones derived in the previous section have been reported by B. Bruhn et al \cite{Bruhn1,Bruhn2}. Recently, numerical simulations of a discrete self-consistent wave-particle model describing an ensemble of particles coupled to a monochromatic wave have revealed a new phenomenon, a phase transition associated with the Landau damping regime. This latter finding adds particular proof to the result presented here.

An interesting continuation of the present study would be the consideration of large-scale hydrodynamic fluctuations and their influence on the dynamics of perturbations. This we plan to complete in a future publication.

\nocite{*}
\bibliography{aipsamp}

\end{document}